\documentclass[aps,pre,twocolumn,superscriptaddress,longbibliography]{revtex4-2}

\usepackage{graphicx}
\usepackage{dcolumn}
\usepackage{bm}
\usepackage{color}
\usepackage[colorlinks=true,allcolors=blue,breaklinks=true]{hyperref}
\usepackage{amsmath} 
\usepackage{siunitx}
\usepackage{multirow}
\usepackage{ctable}

\begin{document}

\title{Evolution of Adaptive Force Chains in Reconfigurable Granular Metamaterials}

\author{Sven Witthaus}
\thanks{These authors contributed equally to this work}
\affiliation{Department of Mechanical Engineering, Yale University, New Haven, Connecticut 06520, USA}

\author{Atoosa Parsa}
\thanks{These authors contributed equally to this work}
\affiliation{Department of Biology, Tufts University, 200 Boston Ave., Medford, MA 02155, USA}

\author{Dong Wang}
\affiliation{Department of Mechanical Engineering, Yale University, New Haven, Connecticut 06520, USA}

\author{Nidhi Pashine}
\affiliation{Department of Mechanical Engineering, Yale University, New Haven, Connecticut 06520, USA}

\author{Jerry Zhang}
\affiliation{Department of Mechanical Engineering, Yale University, New Haven, Connecticut 06520, USA}

\author{Arthur K. MacKeith}
\affiliation{Department of Mechanical Engineering, Yale University, New Haven, Connecticut 06520, USA}

\author{Mark D. Shattuck}
\affiliation{Benjamin Levich Institute and Physics Department, The City College of New York, NY, New York 10031, USA}

\author{Josh Bongard}
\affiliation{Department of Computer Science,  University of Vermont, Innovation 428, Burlington, VT 05405, USA}

\author{Corey S. O'Hern}
\affiliation{Department of Mechanical Engineering, Yale University, New Haven, Connecticut 06520, USA}
\affiliation{Department of Physics, Yale University, New Haven, Connecticut 06520, USA}

\author{Rebecca Kramer-Bottiglio}
\affiliation{Department of Mechanical Engineering, Yale University, New Haven, Connecticut 06520, USA}

\date{\today}

\begin{abstract}
Under an externally applied load, granular packings form force chains that depend on the contact network and moduli of the grains. In this work, we investigate packings of variable modulus (VM) particles, where we can direct force chains by changing the Young's modulus of individual particles within the packing on demand. Each VM particle is made of a silicone shell that encapsulates a core made of a low-melting-point metallic alloy (Field’s metal). By sending an electric current through a co-located copper heater, the Field’s metal internal to each particle can be melted via Joule heating, which softens the particle. As the particle cools to room temperature, the alloy solidifies and the particle recovers its original modulus. To optimize the mechanical response of granular packings containing both soft and stiff particles, we employ an evolutionary algorithm coupled with discrete element method simulations to predict the patterns of particle moduli that will yield specific force outputs on the assembly boundaries. The predicted patterns of particle moduli from the simulations were realized in experiments using 2D assemblies of VM particles and the force outputs on the assembly boundaries were measured using photoelastic techniques. These studies represent a step towards making robotic granular metamaterials that can dynamically adapt their mechanical properties in response to different environmental conditions or perform specific tasks on demand.
\end{abstract}

\maketitle

\section{INTRODUCTION}
\label{sec:intro}
 
\begin{figure*}[!htp]
\centering
\includegraphics[width=\textwidth]{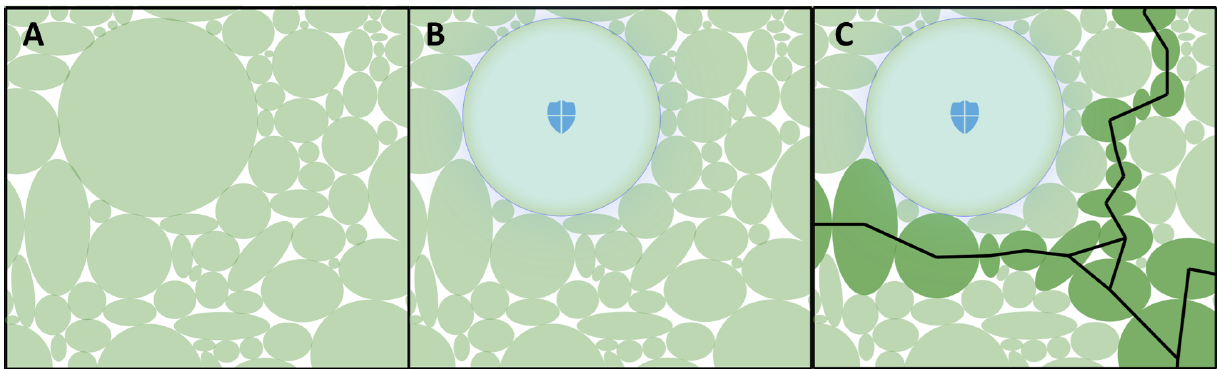}
\caption{Schematic of an envisioned granular packing made of variable modulus (VM) particles. (A) An original jammed granular packing. (B) We wish to shield a target particle in the middle of the packing from applied forces. For example, a damaged particle might need to be shielded if it cannot withstand a given load. (C) By changing the Young's modulus of some of the particles, we can redirect the force network (indicated by solid black lines) to avoid the target particle, allowing the system to withstand larger loads. Soft and stiff particles are depicted in light and dark green, respectively.}
\label{fig1}
\end{figure*}

Mechanical metamaterials process mechanical inputs, such as forces, pressures, or waves, to achieve programmable stiffness, shape transformations, and force propagation~\cite{Kirigamishapetransform, reviewauxetic, fabricmetamaterial}. Many current mechanical metamaterial approaches employ continuum solids or linkages/mechanisms with a fixed structure and therefore demonstrate only fixed responses. We are interested in mechanical metamaterials that can exhibit increased dynamic plasticity, enabling adaptation to different environmental inputs or task demands by reconfiguring their physical structure. Granular metamaterials---consisting of discrete, macroscale particles---offer an advantageous platform for such dynamic programmability, as individual particle properties can be tuned to achieve different bulk responses~\cite{jenett2020discretely}. For example, one can imagine a future granular metamaterial that uses changes in individual particle properties to route forces around a damaged region of the material thereby maintaining its function (Fig.~\ref{fig1}). 

\noindent

Existing granular metamaterials most often incorporate ordered particle assemblies, as it is difficult to predict the material properties of disordered granular assemblies. Ordered granular metamaterials have been used for vibration mitigation~\cite{gantzounis2013granular, leonard2014wave, bonanomi2015wave, nejadsadeghi2019frequency, kim2019wave}, acoustic switches~\cite{li2014granular, wu2019active}, energy absorption~\cite{fu2019programmable}, elastic modulus and density tuning~\cite{porter2015granular, nejadsadeghi2019axially}, and non-reciprocal behaviors~\cite{merkel2019ultrasonic}. Most of these prior studies have confined their approach to granular systems with inert, rigid grains and fixed elastic moduli. In this work, we expand the parameter space by including variable modulus particles, thereby enabling access to a wider set of possible material responses.

In previous studies, we developed discrete element method (DEM) simulations to study the vibrational response of jammed packings composed of binary mixtures of spherical particles with different masses. For example, in one recent study~\cite{wu2019active}, we showed that 2D and 3D mixed-mass granular assemblies can be used to either transmit or block vibrations with particular frequencies. Expanding on that result, we coupled evolutionary algorithms (EA) with DEM simulations to create more complex logic gates~\cite{parsa2022evolution, parsa2022evolving, parsa2023universal}. Using similar granular systems, in this work we seek to control the transmission of force chains, in both simulations and experiments.

In general, granular packings are collections of macroscopic particles that are densely packed, forming contact networks through which interparticle forces are transmitted. It is well-known that a large fraction of the external load applied to granular packings is often carried by a small number of particles~\cite{herrmann2014physics, Behringer_og, Behringer1999, Howell1999, Utter2004, Majmudar2005ContactFM}, resulting in the emergence of stable force chains~\cite{Jaeger1999}. Due to the large number of degrees of freedom, nonlinear response, and out-of-equilibrium behavior of granular systems, force chain dynamics are challenging to predict analytically. However, there have been many computational studies of force chain networks,  such as contact dynamics simulations~\cite{Radjai1996}, DEM simulations~\cite{Luding1997, Peters2005}, random matrix theory, and statistical mechanics-based models~\cite{Edwards2003}, as well as experimental studies of force distributions~\cite{clark2015prl, papadopoulos2016pre, wang2018prl}. Previous studies have emphasized that individual particle properties, such as the elastic moduli and density, influence how forces propagate through granular assemblies~\cite{CAMPBELL2006208}. We will exploit this relationship by switching the elastic moduli of individual particles to achieve specific force outputs from selected particles.

We realize granular metamaterials with adaptable force chains using an assembly of variable modulus (VM) particles. While there are many potential approaches to particle modulus modulation, we adapt the approach previously used by Pashine,~\textit{et al.}~\cite{pashine2023reprogrammable} to make variable stiffness bonds in an allosteric metamaterial. Our VM particles are fabricated by incorporating a low-melting point alloy (Field’s metal) in soft silicone shells. A VM particle possesses a larger modulus when the Field’s metal core is solid (at low temperature) and a smaller modulus when the core is liquid (at high temperature). 

Using DEM simulations combined with evolutionary algorithms, we identify particle configurations that optimize specific force outputs on the assembly boundaries. We then construct these particle configurations in experiments and show that we can adapt the packing's force chain networks on demand by changing the modulus of individual particles. Finally, we implement a multi-objective optimization pipeline that seeks to maximize specific force outputs and minimize power consumption assuming that the particles consume more power when switching from large to small moduli or {\it vice versa}. In all cases that we consider, the numerical simulations and experiments show consistent results for the output forces. Overall, this work represents a step towards the development of dynamic granular metamaterials that can adapt their properties in response to changing environments or task demands. 

\section{Methods}
\begin{figure*}
\centering
\includegraphics[width=1\textwidth]{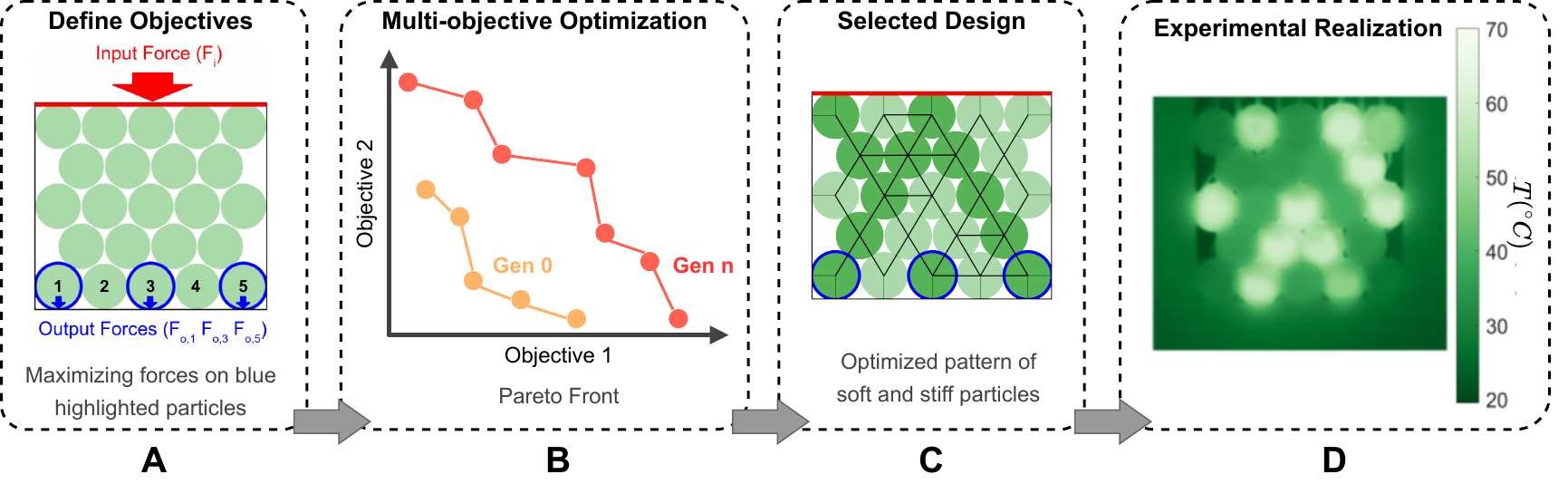}
\caption{Description of the pipeline to generate granular packings with specified force outputs. (A) We start with DEM simulations that can generate force distributions that mimic those of the VM particle packings in experiments. We apply an input force $F_i$ to the top boundary of the packing and measure output forces exerted on the bottom boundary from particles (blue outlines) in the bottom row. We then define optimization objectives to target specific force outputs subject to hardware constraints. (B) Using multi-objective optimization, the evolutionary algorithm identifies a set of particle configurations that best satisfies the objectives. These solutions represent the Pareto front, where each solution represents tradeoffs between the objectives. The algorithm starts with a set of randomly generated configurations in the first generation (Gen 0). During the optimization process, particle configurations that better satisfy the objectives are replaced in the solution set until the stopping criteria are reached (Gen n). (C) An example optimal particle configuration from the Pareto front. Stiff and soft grains are shaded dark and light green, respectively. Black lines indicate inter-particle contacts with the line thickness proportional to the force magnitude. Output forces are maximized for the odd-numbered particles with blue outlines. (D) We can then realize the selected granular packing in experiments and compare the force outputs to those predicted by the DEM simulations. The color scale indicates the particle temperature $T$, which in turn determines the particle moduli (particles with $T > 62^\circ$C are soft and particles with $T < 62^\circ$C are stiff).}\label{fig2} 
\end{figure*}

To controllably modify the force chain network of a packing, we create a pipeline that takes a set of objectives (\textit{e.g.}, maximize the force between a specific particle and the boundary) and hardware constraints as inputs, and outputs the configuration of large- and small-modulus particles that best achieve the objectives. 
Our pipeline has several steps. First, we characterize the contact mechanics and interparticle force law of the VM particles. Second, we develop DEM simulations to represent the physical properties of the granular packings. Finally, we combine DEM simulations and evolutionary algorithms to identify the grain configurations that achieve the objectives (Fig.~\ref{fig2}).

The inverse design problem---designing particle configurations to match a desired force chain output---is an arduous task. Without a systematic methodology, we would have to evaluate an exponentially large number of configurations to decide which configuration gives the desired force output. While it is theoretically feasible to find packings for complex objectives analytically, this task is nontrivial given our nonlinear force model and inability to control the interparticle contacts that occur in granular packings. Therefore, we use DEM simulations coupled with an evolutionary algorithm to identify the optimal grain configurations. In particular, we implement a multi-objective optimization algorithm to search for solutions that satisfy the objectives, subject to experimental constraints. We can then make the grain configurations obtained from the optimization pipeline in experiments and evaluate their force networks, ensuring that the objectives have been met. 

We focus on granular packings composed of monodisperse cylindrical particles (with diameter $D$) confined between two flat walls, such that their cylindrical axes are perpendicular to the walls, and arranged on a 2D triangular lattice. The experimental system is modeled in 2D as collection of monodipserse disks as shown in Fig.~\ref{fig2}A. The packing consists of $N_r = 5$ rows, each with either $4$ or $5$ particles per row, yielding a total number of particles, $N = 23$. We index each particle $p$ between $1$ and $N$, increasing from left to right and from bottom to top in the packing, as shown in Fig.~\ref{fig2}A. For a specific configuration denoted as $C$, we assign each grain with an elastic modulus chosen from one of two values: soft $k_{soft}$ or stiff $k_{stiff}$ with $k_{soft} < k_{stiff}$. We define the input force $F_i$ as the total force applied downwards on the top boundary of the system (Fig.~\ref{fig2}A.) After applying $F_i$, we measure the output forces exerted on the boundary from the $p$th particle in the bottom row, $F_{o,p}$. 

In the remainder of this section, we describe in detail the characterization of the Young's modulus of the particles and interparticle contact force law, the experimental setup, the DEM simulations, and the optimization approaches.

\subsection{Experimental Design}

\subsubsection{Variable Modulus Particles}
\begin{figure*}[!htbp]
\centering
\includegraphics[width=\linewidth]{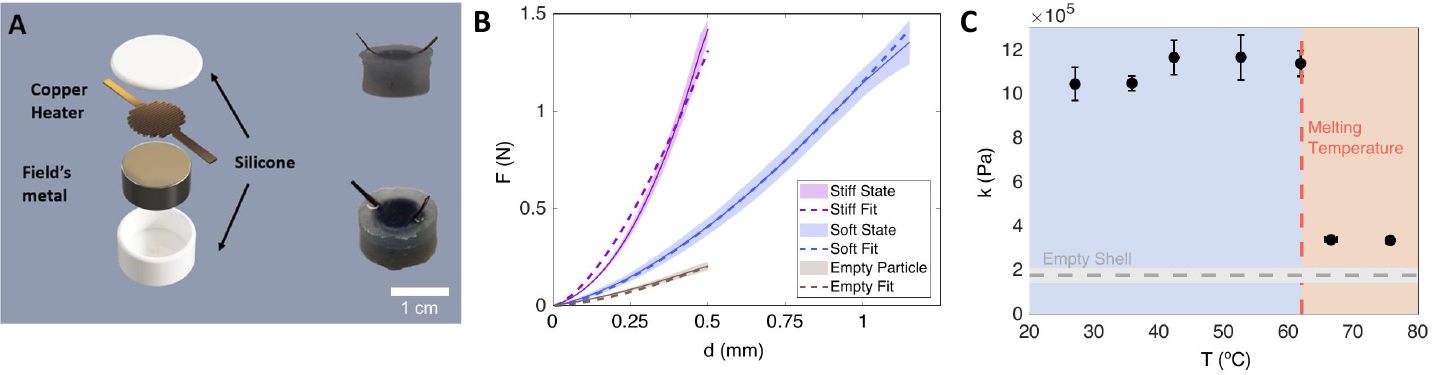}
\caption{Variable modulus (VM) particle characterization. (A) Illustration of the components of a VM particle (left) and an image of a fully assembled VM particle (right). (B) Force $F$ on VM particle from the two walls in the Instron plotted versus the symmetric displacement $d$ of the walls (relative to the location of the walls at contact) in the soft (blue solid lines) and stiff (magenta solid lines) states. The dashed lines indicate fits to Eq.~\ref{eq:hertzian} for the Hertzian contact law. Data for a particle with the same silicone geometry, but without Field's metal inside, is labeled as an ``empty particle.'' Shaded areas represent $\pm$1~standard deviation from six trials. (C) Particle modulus $k$ plotted as a function of temperature $T$. The melting temperature of the Field's metal is highlighted by the vertical dashed line. $k$ for the ``empty particle'' is also indicated by the horizontal dashed line.}\label{fig3}
\end{figure*}

We make the cores of the VM particles using Field's metal, an alloy of bismuth, indium, and tin with a melting temperature of $T_0 = 62^{\circ}$C. We encapsulate a solid cylinder of Field's metal inside a soft silicone shell made from Smooth-on Dragonskin\texttrademark~10 (Fig.~\ref{fig3}A). We place a small copper heater inside the shell that allows us to melt the Field's metal via Joule heating. The high-resistance copper heater is separated from the Field's metal by a thin layer of silicone to avoid shorting the heater. By running a current ($\approx 1 \text{A}$) through the resistive heater we induce Joule heating in the particles. This heating melts the Field's metal inside at around \SI{3.5}W, which changes its Young's modulus. While solid, Field's metal has a Young's modulus of $\SI{9.25}{\giga \pascal} $~\cite{Shan_2013}, which reduces to near zero at the melting temperature.

Next, we describe measurements of the Young's modulus of the VM particles. We compress single VM particles between two steel walls using a material testing system (Instron\texttrademark~$3365$). We find that the relationship between the symmetric displacement $d$ of the walls (relative to their location at contact with the particle) and the compressive force $F$ is best fit by assuming a Hertzian contact law~\cite{Hertzian_cylinder}: 
\begin{equation}
    F = \frac{4}{3} E^* R^{*\frac{1}{2}} d^{\frac{3}{2}},
    \label{eq:hertzian}
\end{equation}
where 
\begin{equation}
    \frac{1}{E^*} = \frac{1 - \nu_s^2}{E_s} + \frac{1 - \nu^2}{E},
    \label{eq:E-eff}
\end{equation}
$E_s$ and $E$ are the Young's moduli of steel and the VM particle, $\nu_s$ and $\nu$ give the Poisson ratios of steel and the VM particle,
\begin{equation}
    \frac{1}{R^*} = \frac{1}{R_s} + \frac{1}{R},
    \label{eq:R-eff}
\end{equation}
and $R_s$ and $R$ are the radii of the steel walls and the VM particle. We assume that $E_s \rightarrow \infty$ and $R_s \rightarrow \infty$. Note that Eq.~\ref{eq:hertzian} gives the contact force law between two elastic spheres, not between two elastic cylinders with parallel cylindrical axes. The force law between the cylinders in the experiments can mimic the force law between contacting spheres when the cylinders have shape imperfections and the cylindrical axes are not parallel~\cite{franklin2016handbook, zhao2022prx}. The force $F$ versus displacement $d$ is shown in Fig.~\ref{fig3}B for the silicone shell alone, the VM particle when heated (soft state), and the VM particle at room temperature (stiff state). Fitting the force versus the displacement curve yields the VM particle modulus:
\begin{equation}
    k = \frac{E}{1 - \nu^2}.
    \label{eq-stiffness}
\end{equation}
In Fig.~\ref{fig3}C, we plot the dependence of $k$ on the temperature $T$ of the Field's metal. The particle modulus decreases rapidly when $T > T_0$. We find $k_{soft} \sim 0.34$~MPa for the soft state ($T>T_0$) and $k_{stiff} \sim 1$~MPa for the stiff state ($T < T_0$). Thus, the VM particle design possesses a three-fold change in modulus.  $k_{soft}$ is roughly $1.5$ times the modulus of the empty silicone shell. For the empty silicone shell, we report $k \approx 0.2$~MPa, which is below that for a solid silicone material ($\sim 0.53$~MPa). Also note that since the Field's metal becomes a liquid when heated above the melting temperature, an isolated Field's metal particle can in principle yield an infinite-fold modulus change between the liquid and solid states.  The three-fold change for the VM particle modulus highlights the substantial contribution of the outer silicone shell to the VM particle modulus. Since the silicone shell encapsulates the metal core, compressive strains predominantly affect the silicone rather than the metal, reducing the modulus ratio between the stiff and soft states compared to the case where only the metal is compressed in the absence of the silicone shell.

\subsubsection{Particle Assembly}
\label{sect212}
 
We assemble the VM particles into a triangular lattice where each particle can be actuated (heated) individually using a microcontroller (Arduino\texttrademark). However, heat from actuated particles eventually dissipates into the neighboring non-actuated particles, causing thermal crosstalk. To mitigate thermal crosstalk, we established limits on the actuation power of each VS particle (see ESI S1). 

We apply a uniform load $F_i$ to the top layer of particles using the same materials testing system used to characterize the VM particles. The force propagates through the packing to create a particular force chain network that is a function of the positions of the particles and their individual moduli. Based on the locations of the stiff and soft particles, the load is transmitted differently to the bottom row of particles, exerting different amounts of force at various locations on the bottom boundary.

The assembly boundaries are made of a photoelastic material (ClearFlex\texttrademark~95) placed between two circular-crossed polarizers. We measure the output forces $F_{o,p}$ on the bottom boundary by analyzing the stress-induced birefringence patterns \cite{deJong2006, Takaophoto, Utter2004}. When the boundary is not under stress, no light passes through the polarizers. Any stress on the boundary rotates the direction of the incoming electric field, allowing light to pass through the polarizer and appear as a fringe pattern. Because different wavelengths of light will form different patterns on the photoelastic boundary, we use a filter to only observe one wavelength of light ($\lambda =530$~nm). 

To measure $F_{o,p}$, we consider the intensity $I_{out}$ that occurs through the application of a point force $F_{o,p}$ on an elastic half-plane through a polariscope~\cite{GranularShattuck, flamant1892repartition}. The light intensity $I_{out}$ at any given point $(x, y)$ within the half-plane is
\begin{equation}
    I_{out} = I_0^2 \text{sin}^2 \frac{\pi t K}{\lambda} \left (\sigma_1-\sigma_2\right) ,
  \label{eq2}
\end{equation}
{\noindent}where
$t$ is the thickness of the material, $I_0$ is the maximum intensity, $K$ is the material-dependent stress-optic coefficient, and $\sigma_1-\sigma_2$ is the principle stress difference, which in our system corresponds to the normal stress measured radially outward from the particle-wall contact. 
Therefore, 

\begin{equation}
\sigma_1 - \sigma_2  = \Big| \frac{2}{\pi r} F_{o,p} (\text{cos} \phi \text{sin} \theta + \text{sin} \phi \text{cos} \theta) \Big| , 
\label{eq3}
\end{equation}
{\noindent}where $r$ is the distance from the point force to the point $(x, y)$, $\theta$ is the angle of the vector pointing from the point force to $(x, y)$ relative to the x-axis (parallel to the bottom boundary), and $\phi$ is the angle that determines the components of the force in the $x$-$y$ plane (Fig.~\ref{fig4}A.)

Using Eqs.~\ref{eq2} and~\ref{eq3}, we can construct an image for a given force on a half-plane. By comparing this image to the experimentally obtained image, we can fit for $2tKF_{o,p}/\lambda$ using minimum chi-square estimation~\cite{GranularShattuck}. This fit yields $F_{o,p}$ to within a proportionality factor, $2tK/\lambda$. However, we can eliminate the proportionality constant by summing all of the forces on the bottom boundary: $\sum_{p=1}^{5} F_{o,p} = F_i$. See ESI$\dag$ S2 for more details.

\subsection{DEM simulations}

We employ discrete element method (DEM) simulations in two dimensions that mimic the experimental setup. We assemble a collection of monodisperse, frictionless disks into a triangular lattice (see Fig.~\ref{fig2}A). We apply a constant downward force $F_i/(k_{soft}D^2) = 0.01$ to the top boundary. Each particle interacts with its neighbors and the walls, assuming purely repulsive Hertzian interactions, as in Eq.~\ref{eq:hertzian}. We integrate Newton's equations of motion for each disk with damping forces proportional to the particle velocities using a modified velocity-Verlet integration scheme until the net force on all disks $N^{-1} |\sum_p {\vec F}_p|/(k_{soft} D^2) <10^{-10}$. Determining which disks are in contact allows us to construct the network of interparticle contacts, as shown in Fig.~\ref{fig2}C. In ESI S3, we show that $F_{o,p}$ is proportional to $F_i$ over a broad range of $F_i$. Hence, the specific choice of $F_i$ in the DEM simulations does not impact the optimal solutions predicted by the evolutionary algorithm, as long as $F_i$ is sufficiently small.

We note that there are some differences between the DEM simulations and the experiments. For example, the DEM simulations do not include particle-substrate and inter-particle friction. We mitigate the effects of friction by applying a thin layer of cornstarch over the surfaces of each VS particle in the experiments, thus decreasing the effective friction in the experiments. In addition, the particle sizes are slightly polydisperse ($< 2\%$) in the experiments, while the simulations assume monodisperse disks. Despite these differences, we find strong qualitative agreement between the results of the DEM simulations and the experiments.  

\subsection{Optimization Setup}
To solve the inverse design problem of finding particle configurations that achieve a desired output force pattern, we employ evolutionary algorithms (EAs) in our design pipeline (Fig.~\ref{fig2}). EAs are a class of population-based gradient-free global optimization methods inspired by concepts from biology, genetics, and natural evolution \cite{Goldberg1989}. In the general case, the algorithm starts with a set of randomly generated solutions (called a \textit{population}). In each subsequent optimization step (\textit{i.e.}, \textit{generation}), possible solutions are evaluated using a fitness function. The solutions with the highest fitness are chosen to survive to the next generation and produce alternate solutions through a process of mutation and selection that can potentially create solutions with higher fitness. EAs have been useful in myriad problems in materials science~\cite{miskin2014evolving}, robotics~\cite{shah2021shape}, and data retrieval applications~\cite{chiong2012variants}.

In this study, to find an optimal configuration of soft and stiff particles to achieve a desired force output, we utilize a particular EA called Age-Fitness Pareto Optimization (AFPO) \cite{schmidt2010age}. AFPO is a multi-objective evolutionary algorithm that balances the diversity of solutions with their fitness values, allowing for an extensive search in configuration space without converging to local optima in a rugged fitness landscape. AFPO converges to a set of optimal solutions that is often referred to as the Pareto front. No solution is better than the others on the Pareto front (and each solution optimizes the multiple objectives differently) unless additional trade-offs are specified. To select the best solution, one can rank Pareto-front solutions with methods used for multiple criteria decision-making, such as the Technique for Order of Preference by Similarity to Ideal Solution (TOPSIS) method~\cite{hwang1981methods}. TOPSIS implements trade-offs between objectives that are opposing to each other by assigning a weight vector $W$ that indicates how much each given objective should be weighted while traversing the Pareto front. TOPSIS then chooses the solution with the shortest distance from the ideal solution and the greatest distance from the negative ideal (worst) solution.  Note that we can also use AFPO for single-objective optimization, in which case we do not need to consider the Pareto front.

\begin{table*}[!t]
\small
  \caption{\ Details of the AFPO algorithm for the single- and multi-objective optimizations carried out in this study.}
  \label{tbl:parameters}
  \begin{tabular*}{0.9\linewidth}{@{\extracolsep}l @{\hspace{.3\tabcolsep}} | l @{\hspace{.3\tabcolsep}} l}
  
    \specialrule{.1em}{0.2em}{0.2em} 
    
    \multicolumn{1}{c|}{\textbf{Parameter}} & \multicolumn{2}{c}{\textbf{Value}} \\
    
    \specialrule{.1em}{0.2em}{0.2em} 

    Genome Representation ($C$) & \multicolumn{2}{l}{binary string $C = [x_{1}, x_{2}, ..., x_p, ..., x_{23}], x_p \in \{0, 1\}$} \\[0.15cm]

    \specialrule{.01em}{0em}{0.2em} 
    
    Phenotype ($\mathcal{K}$) & \multicolumn{2}{l}{$\mathcal{K} = [k_{1}, k_{2}, ..., k_{p}, ..., k_{23}], k_p = x_p k_{soft}+(1-x_p)k_{stiff}, k_{stiff}/k_{soft}=3.0$} \\[0.15cm]

    \specialrule{.01em}{0em}{0.2em} 
    
    Variation Operator & \multicolumn{2}{c}{bit-flip mutation} \\[0.15cm]

    \specialrule{.01em}{0em}{0.2em} 
    
    Mutation Probability & \multicolumn{2}{c}{$0.05$} \\[0.15cm]

    \specialrule{.01em}{0em}{0.2em}  
    
    \multirow{5}{*}{Objective Functions} & \textit{Feasibility Check} & ${\mathcal O_0}={F_{o,3}}/{F_i}$ \\[0.1cm] & \textit{Case I} & $\mathcal{O}_1 = {\sqrt[3]{F_{o,1} \; F_{o,3} \; F_{o,5}}}/{F_i}$ \\[0.1cm] & \textit{Case II} & $\mathcal{O}_2 ={\sqrt{F_{o,2} \; F_{o,4}}}/{F_i}$ \\[0.1cm] & \textit{Case III} & $\mathcal{O}_2$ and $\mathcal{O}_3 = N_{C_1^* \rightarrow C_2}$ \\[0.1cm] & \textit{Case IV} & $\mathcal{O}_1$ and $\mathcal{O}_4 = N_{C_2^* \rightarrow C_1}$ \\[0.15cm]

    \specialrule{.01em}{0em}{0.2em}

    \multirow{2}{*}{Population Size ($P$)} & \textit{Feasibility Check} & \multicolumn{1}{c}{$P = 50$} \\[0.1cm] & \textit{Cases I - IV} & \multicolumn{1}{c}{$P = 100$} \\[0.15cm]

    \specialrule{.01em}{0em}{0.2em} 
    
    \multirow{2}{*}{Number of Generations ($G$)} & \textit{Feasibility Check} & \multicolumn{1}{c}{$G = 250$} \\[0.1cm] & \textit{Cases I - IV} & \multicolumn{1}{c}{$G = 500$} \\

    \specialrule{.01em}{0em}{0.2em} 
    
    Initialization & \multicolumn{2}{l}{$\{C^n\}, \; n=1...P$, for each $n$, $[x_p] = \mathcal{U}(0, 1)$} \\[0.15cm]

    
    
    \specialrule{.1em}{0.2em}{0.2em} 
    
  \end{tabular*}
\end{table*}

The AFPO algorithm has several components including a genome representation, a variation operator, and a fitness function. In this study, we used a direct encoding scheme for the genome, where the genotype is a binary string ($0$ for $k_{soft}$ and $1$ for $k_{stiff}$) that represents the modulus values for all of the particles. The initial genotype is a string of $N$ random bits, $\mathcal{U}(0, 1)$.  The variation operator is a bit-flip mutation with a given mutation probability. The bit-flip mutation changes the modulus of a particle from $k_{soft}$ to $k_{stiff}$ or \textit{vice versa}. We consider both single- and multi-objective fitness functions. Additional details concerning the implementation of the AFPO algorithm are provided in Table~\ref{tbl:parameters}.

\begin{figure*}[!ht]
\centering
\includegraphics[width=\textwidth]{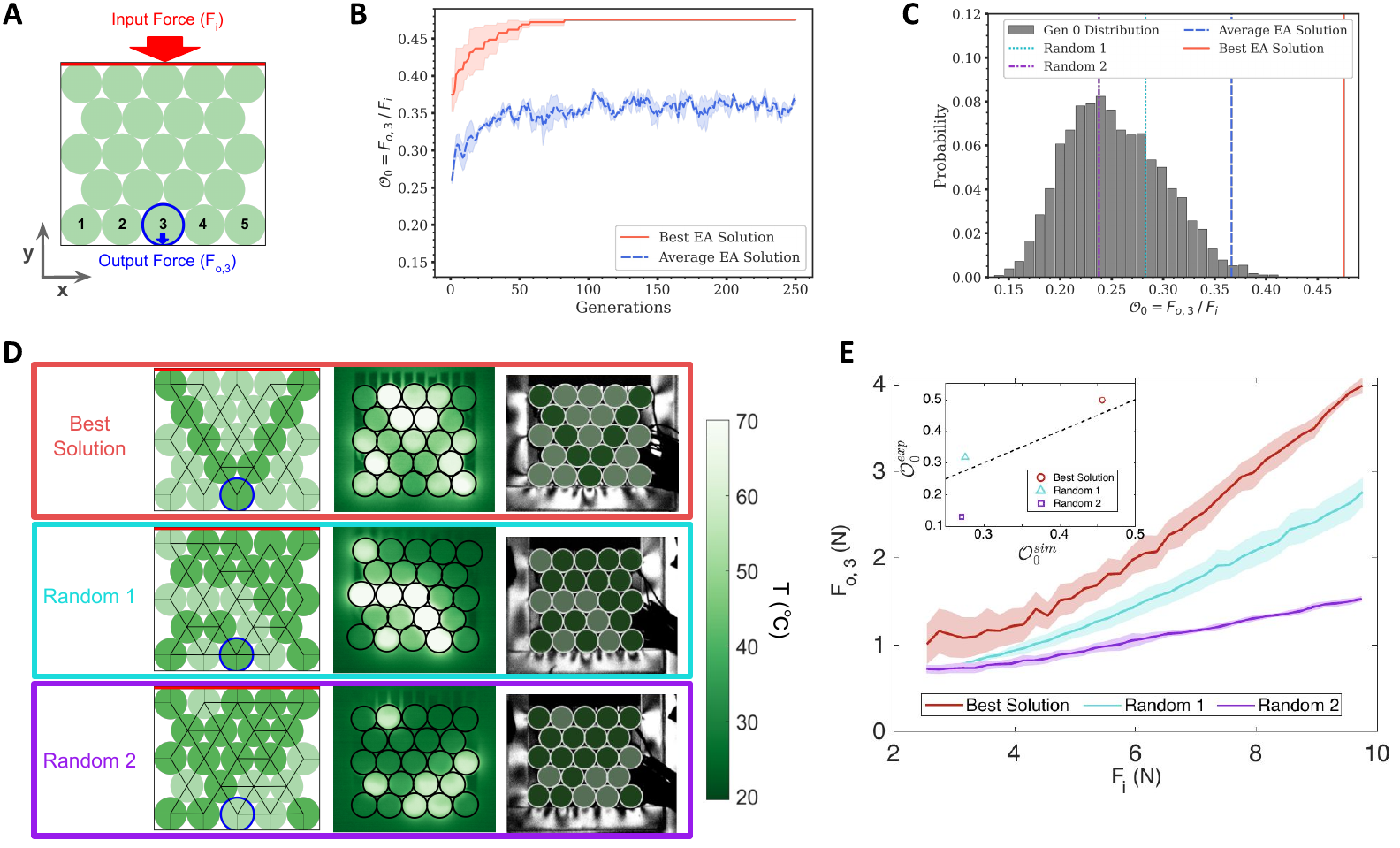}
\caption{Assessment of the optimization of the VM particle assembly. (A) Schematic of the particle assembly. The optimization objective $\mathcal{O}_0$ is to maximize the force output $F_{o,3}$ of the middle particle in the bottom row given an input force applied to the top boundary $F_i$. 
(B) $\mathcal{O}_0$ for the best solution and averaged over the whole population, plotted as a function of the number of generations. The results are averaged over three independent trials with random initialization. (C) The probability distribution of $\mathcal{O}_0$ for $5000$ randomly generated configurations with vertical lines that indicate $\mathcal{O}_0$ for the average (blue dashed line) and best solutions (red solid line) at $G=250$. We also show $\mathcal{O}_0$ for two random configurations (Random 1: cyan dotted line and Random 2: magenta dashed-dotted line) pictured in panel D. (D) The first column shows three configurations with the predicted inter-particle contact networks from the DEM simulations. The soft and stiff particles are shown in light and dark shades of green, respectively. The best solution found by EA is shown in the first row, along with two random configurations (Random 1 and Random 2) on the second and third rows. The second column shows thermal images of the experimental packings with temperature increasing from dark green to white. The third column illustrates the birefringence patterns on the photoelastic boundaries. The black cords emanating from the bottom right of the packings provide electrical current to each particle. (E) $F_{o,3}$ plotted as a function of $F_i$ obtained from experiments for the same configurations in (D). Shaded areas represent $\pm$1~standard deviation from three independent trials. In the inset to (E), we plot the objective function from the experiments ${\cal O}_0^{exp}$ versus that in the the DEM simulations ${\cal O}_0^{sim}$, where the dashed line indicates $\mathcal{O}_0^{exp} = \mathcal{O}_0^{sim}$.}\label{fig4}
\end{figure*}

We explore five optimization objectives as outlined in Table~\ref{tbl:parameters}. In the \textit{Feasibility Check}, we consider a single-objective optimization problem to maximize the force output from a single particle on the bottom row (Sec.~\ref{sect31}). We then consider two other single-objective optimization problems to further test the efficacy of our optimization pipeline (\textit{Cases I} and \textit{II}). In these two cases, the objective function is defined so that the force output from the odd- or even-numbered particles on the bottom boundary is maximized (Sec.~\ref{sect321}). In Sect.~\ref{sect322}, we consider a multi-objective optimization problem (\textit{Cases III} and \textit{IV}) where we can change from the optimal configuration in \textit{Case I} to that in \textit{Case II} (or vice versa) with an additional hardware constraint that limits the number of changes in the particle modulus as the second objective. The formulation of the objective functions and optimization results for each case are presented in the next section. 

\section{Results}
\label{results}

\subsection{Feasibility Check: Maximizing the output force from a single particle on the bottom row}
\label{sect31}
To verify that the DEM simulations are sufficiently accurate and that the EA generates optimal solutions, we first study a simple case of maximizing the output force exerted by the middle particle in the bottom row on the boundary ($F_{o, 3}$ in Fig.~\ref{fig4}A). For this case, we define the objective function,
\begin{equation}
    {\mathcal O}_0=\frac{F_{o,3}}{F_i}, 
    \label{eq-o}
\end{equation}
and the goal is to find a configuration $C^*$ that maximizes $\mathcal{O}_0$. As shown in Fig.~\ref{fig4}B, using EA, the optimal configuration emerges after ${\sim}100$ generations and has higher $\mathcal{O}_0$ than that generated using a random search with over $5000$ configurations (Fig.~\ref{fig4}C).

\begin{figure*}[!ht]
\centering
\includegraphics[width=\textwidth]{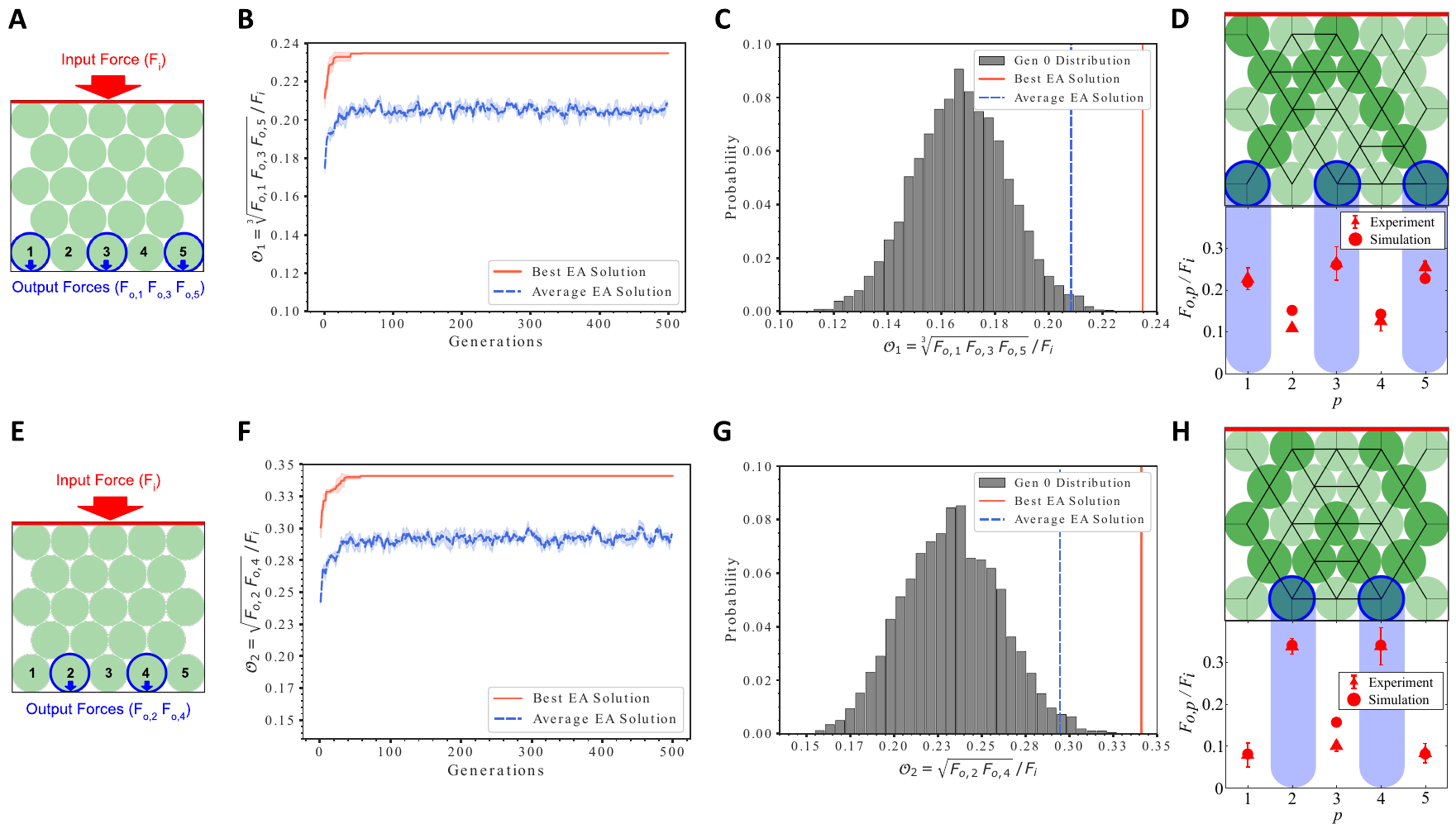}
\caption{Assessment of the optimization for \textit{Cases I} and \textit{II}. (A) For Case I, we seek the optimal solution $C^*_1$ that maximizes the output forces from the odd-numbered particles on the bottom row using the objective function $\mathcal{O}_1$ in Eq.~\ref{eq4}. (B) $\mathcal{O}_1$ for the best solution and averaged over the whole population plotted versus the number of generations. The results are averaged over three independent trials with random initialization. (C) The probability distribution of $\mathcal{O}_1$ for $5000$ randomly generated configurations with vertical lines that indicate $\mathcal{O}_1$ for the average (blue dashed line) and best solution (red solid line) at $G=500$. (D) We show one of the two optimal asymmetric solutions for ${\cal O}_1$, $C^*_{1a}$. The other optimal solution $C^*_{1b}$ is a mirror image of $C^*_{1a}$. We plot $F_{o,p}$ versus the location $p$ of the particles on the bottom row from the DEM simulations and experiments. The blue shading indicates the odd-numbered particles. (E) For Case II, we seek the optimal solution $C^*_2$ that maximizes the output forces from the even-numbered particles on the bottom row using the objective function $\mathcal{O}_2$ in Eq.~\ref{eq5}. (F) $\mathcal{O}_2$ for the best solution and averaged over the whole population plotted versus the number of generations. The results are averaged over three independent trials with random initialization. (G) The probability distribution of $\mathcal{O}_2$ for $5000$ randomly generated configurations with vertical lines that indicate $\mathcal{O}_2$ for the average (blue dashed line) and best solution (red solid line) at $G=500$. (H) $C^*_2$ is the optimal solution for ${\cal O}_2$. We plot $F_{o,p}$ versus $p$ from the DEM simulations and experiments. The blue shading indicates the even-numbered particles.} \label{fig5}
\end{figure*}

In Fig.~\ref{fig4}D, we show that $C^*$ contains two lines of stiff particles stretching from the top corners to the central particle in the bottom row. This solution is intuitive, as larger forces will be carried by the stiffer particles. Thus, arranging stiff particles such that they form a connected path from the top to the bottom boundary without splitting will result in the largest force carried by the central bottom particle. 

To validate the results obtained by the EA, we first construct the three configurations (Best Solution, Random 1, and Random 2) in the experiments (Fig.~\ref{fig4}D) and measure $F_{o, 3}$ as a function of $F_i$ for each configuration.  In Fig.~\ref{fig4}E, we show that the output force $F_{o,3}$ is approximately linear with $F_i$ for $F_i > 4$~N, which is consistent with the results from the DEM simulations in ESI S3. For $F_i < 4$~N, $F_{o,3}$ is nearly constant, which is caused by the low signal-to-noise values for the photoelastic material at small forces. We also note that the top plate provides comparable forces to each particle in the top row (\textit{e.g.}, $\sim 10~N/5$ yields $\sim 2$~N per particle) and these force magnitudes are similar to those used to measure the stiffness of the VM particles ($\sim 1.5$~N, Fig.~\ref{fig3}B). We determine $\mathcal{O}_0$ by calculating the slope of $F_{o,3}$ versus $F_i$ for $F_i > 4$~N in experiments. As shown in Fig.~\ref{fig4}E, the best solution ($\mathcal{O}_0^{exp} \approx 0.5$) substantially outperforms the two random configurations ($\mathcal{O}_0^{exp} \approx 0.32$ and $0.13$ for Random 1 and Random 2, respectively).

We observe quantitative agreement for ${\cal O}_0$ between the experiments and DEM simulations for the Best Solution and Random 1 configurations: $\mathcal{O}_0^{sim} \approx {\cal O}_0^{exp} \approx 0.46$ for the Best Solution and $\mathcal{O}_0^{sim} \approx {\cal O}_0^{exp} \approx 0.28$ for Random 1. However, $\mathcal{O}_0^{sim} \approx 0.27$ for Random 2 is approximately twice that measured in experiments. (Note that this discrepancy will not affect our results since the $\mathcal{O}_0^{sim}$ value for Random 2 is much less than that for the Best Solution.) These results demonstrate our ability to experimentally construct configurations with maximal ${\cal O}_0$ that have been predicted by EA.

\subsection{Maximizing force outputs from multiple particles}
\label{sect321}

We next consider objective functions that yield configurations with maximal output forces from mutliple particles in the bottom row, instead of just one particle as described in Sec.~\ref{sect31}. Specifically, we will consider the following two cases:
\begin{itemize}
    \item In \textit{Case I}, we seek a configuration $C^*_1$ that maximizes the force outputs from the odd-numbered particles on the bottom row (Fig.~\ref{fig5}A.)
    \item In \textit{Case II}, we seek a configuration $C^*_2$ that maximizes the force outputs from the even-numbered particles on the bottom row (Fig.~\ref{fig5}E.)
\end{itemize}

For Case I, we define the objective function,
\begin{equation}
\label{eq4}
    \mathcal{O}_1 = \frac{\sqrt[3]{F_{o,1} \; F_{o,3} \; F_{o,5}}}{F_i},
\end{equation}
which is the geometric mean of the force outputs of the odd-numbered particles on the bottom row. Similarly, for Case II, we define
\begin{equation}
\label{eq5}
    \mathcal{O}_2 = \frac{\sqrt{F_{o,2} \; F_{o,4}}}{F_i},
\end{equation}
which is the geometric mean of the force outputs of the even-numbered particles on the bottom row.  Both ${\cal O}_1$ and ${\cal O}_2$ are normalized by the input force $F_i$. For both Cases I and II, we use the optimization pipeline described in Table~\ref{tbl:parameters}.

Similar to the case of the Feasibility Check, the optimal configuration $C_1^*$ that maximizes $\mathcal{O}_1$ for Case I emerges after $\sim 80$ generations (Fig.~\ref{fig5}B) and has a much larger $\mathcal{O}_1$ than that generated using a random search with over $5000$ configurations (Fig.~\ref{fig5}C). $C^*_1$ possesses lines of stiff particles that extend from the force input on the top row of particles to the odd-numbered particles in the bottom row (Fig.~\ref{fig5}D), similar to the stiffness pattern in $C^*$. 
Note that the stiffness pattern in $C_1^*$ is not symmetric about the $y$-axis, hence there are two solutions: $C_{1a}^*$ shown in Fig.~\ref{fig5}D, and a mirror image of $C^*_{1a}$ about the $y$-axis with different genome representations (Table~\ref{tbl:parameters}), but the same ${\cal O}_1$. In addition, in Fig.~\ref{fig5}D we show that the $F_{o,p} / F_i$ values in the experiments are close to those predicted by the DEM simulations for all particles $p$. In addition, all odd-numbered particles have higher output forces than those for the even-numbered particles on the bottom row, which is consistent with maximizing $\mathcal{O}_1$.

We observe similar results for Cases I and II. We obtain the optimal configuration $C_2^*$ that maximizes $\mathcal{O}_2$ after $\sim 80$ generations (Fig.~\ref{fig5}F). It has a much larger $\mathcal{O}_2$ than that generated using a random search (Fig.~\ref{fig5}G). $C^*_2$ is symmetric about the $y$-axis and the lines of stiff particles are reflected at the side walls, which directs forces from the top plate to the even-numbered particles and avoids the propagation of large forces to the odd-numbered particles (Fig.~\ref{fig5}H). We again find that $F_{o,p} / F_i$ in the experiments are very close to those predicted by the DEM simulations for all $p$, as shown in Fig.~\ref{fig5}H. In $C_2^*$, all even-numbered  particles have higher output forces than those for the odd-numbered particles on the bottom row, which is consistent with maximizing $\mathcal{O}_2$.

\subsection{Optimal solutions with two objectives}
\label{sect322}

Given that there are different force outputs from the particles on the bottom row in $C_1^*$, and $C_2^*$, one can envision an application where the system needs to switch from $C_1^*$ (large force outputs from odd-numbered particles) to $C_2^*$ (large force outputs from even-numbered particles) or {\it vice versa}.  The VM particles provide the capability to switch the packing from one configuration to another, \textit{i.e.}, by local heating, one can change $k$ for each particle from soft to stiff and {\it vice versa}. In contrast, to switch from $C_1^*$ to $C_2^*$ with inert particles, one would need to create $C_1^*$, disassemble it, and then build $C_2^*$. In the movie in the ESI, we show that using VM particles we can change the stiffness network from $C_1^*$ to $C_2^*$ without disassembling the packing by changing the temperatures of the appropriate particles.

\begin{figure}[!t]
\centering
\includegraphics[width=\columnwidth]{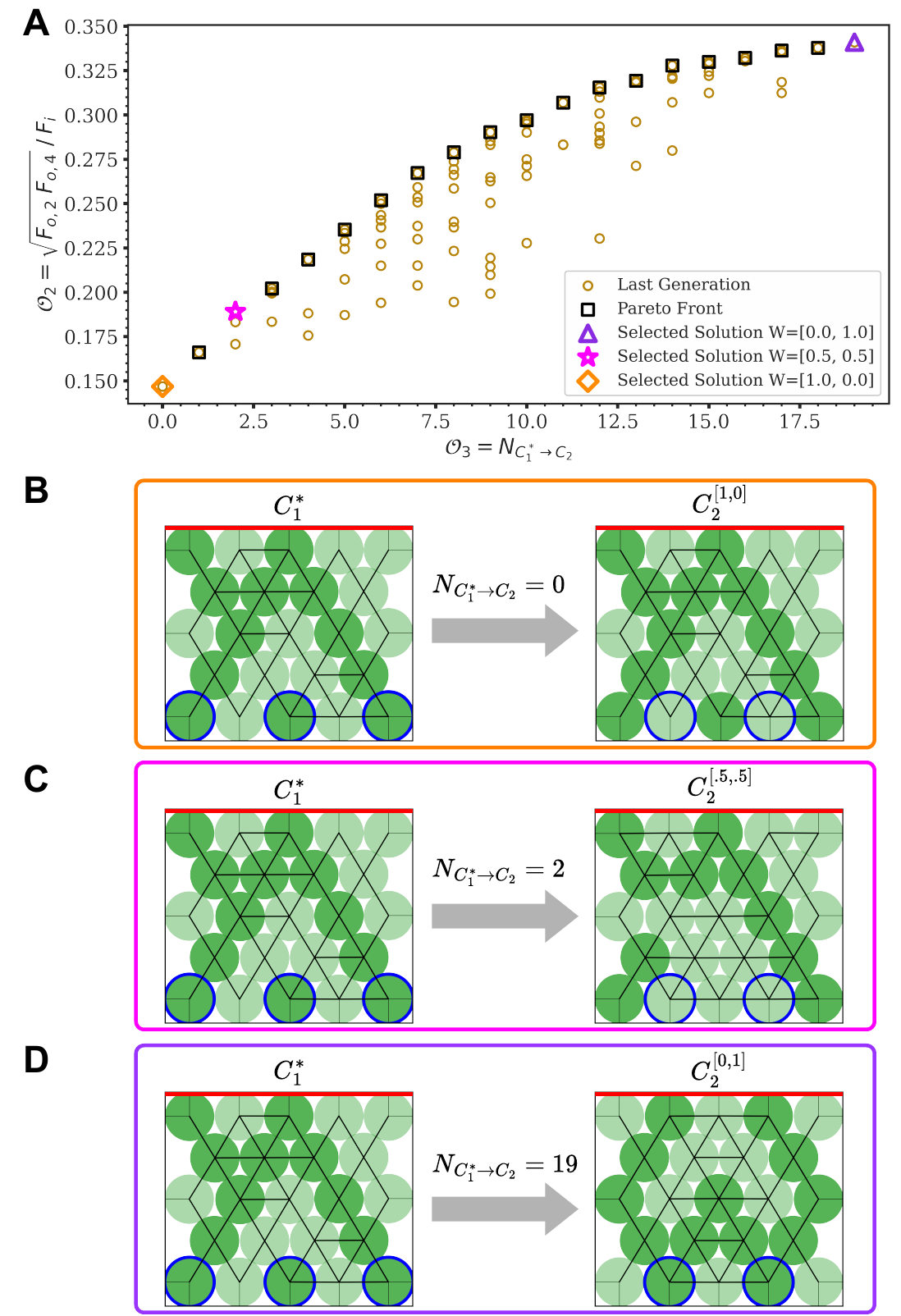}
\caption{Multi-objective optimization to find a configuration that maximizes the output forces on the even-numbered particles while minimizing the number of particle modulus changes starting from $C_1^*$. (A) Solutions in the $\mathcal{O}_3$-$\mathcal{O}_2$ plane at $G=500$ (yellow circles). We also highlight the Pareto front (black squares). The orange diamond, magenta star, and purple triangle indicate the best solutions based on the TOPSIS method with different weight vectors $W$. (B-D) Starting from $C_1^*$ (left), we show the best solution (right) from the Pareto front with a given TOPSIS weight vector $W$: (B) $W=[1, 0]$, $C^{[1,0]}_2$, (C) $W=[0.5, 0.5]$, $C^{[.5,.5]}_2$, and (D) $W=[0, 1]$, $C^{[0,1]}_2$. The first and second elements of $W$ correspond to the weights for $\mathcal{O}_3$ and $\mathcal{O}_2$, respectively.}\label{fig6}
\end{figure}

To change from $C_1^*$ to $C_2^*$ (and {\it vice versa}) requires $N_s=19$ particle modulus changes. However, what if there is a hardware constraint for the VM particles where significant power is consumed to change the modulus, but not to maintain the particle modulus as soft or stiff? In this case, minimization of the power consumption corresponds to minimizing the number of switches in modulus between the two configurations and a multi-objective EA is required to identify the optimal solution.

For \textit{Case III}, we consider the following multi-objective optimization problem. Starting from $C_1^*$, we seek to maximize $\mathcal{O}_2$ and minimize
\begin{equation}
\label{eq6}
    \mathcal{O}_3 = N_{C_1^* \rightarrow C_2} = \sum_{p=1}^N \left(1 - \delta \left(k^{C_1^*, p}, k^{C_2, p} \right) \right),
\end{equation}
where $N_{C_1^* \rightarrow C_2}$ is the number of particle modulus changes from $C_1^*$ to $C_2$, $\delta(x, y) = 1$ when $x = y$ and $\delta(x, y) = 0$ when $x \ne y$, and $k^{C, p}$ is the modulus of the $p$th particle in configuration $C$.

\begin{figure}[!t]
\centering
\includegraphics[width=\columnwidth]{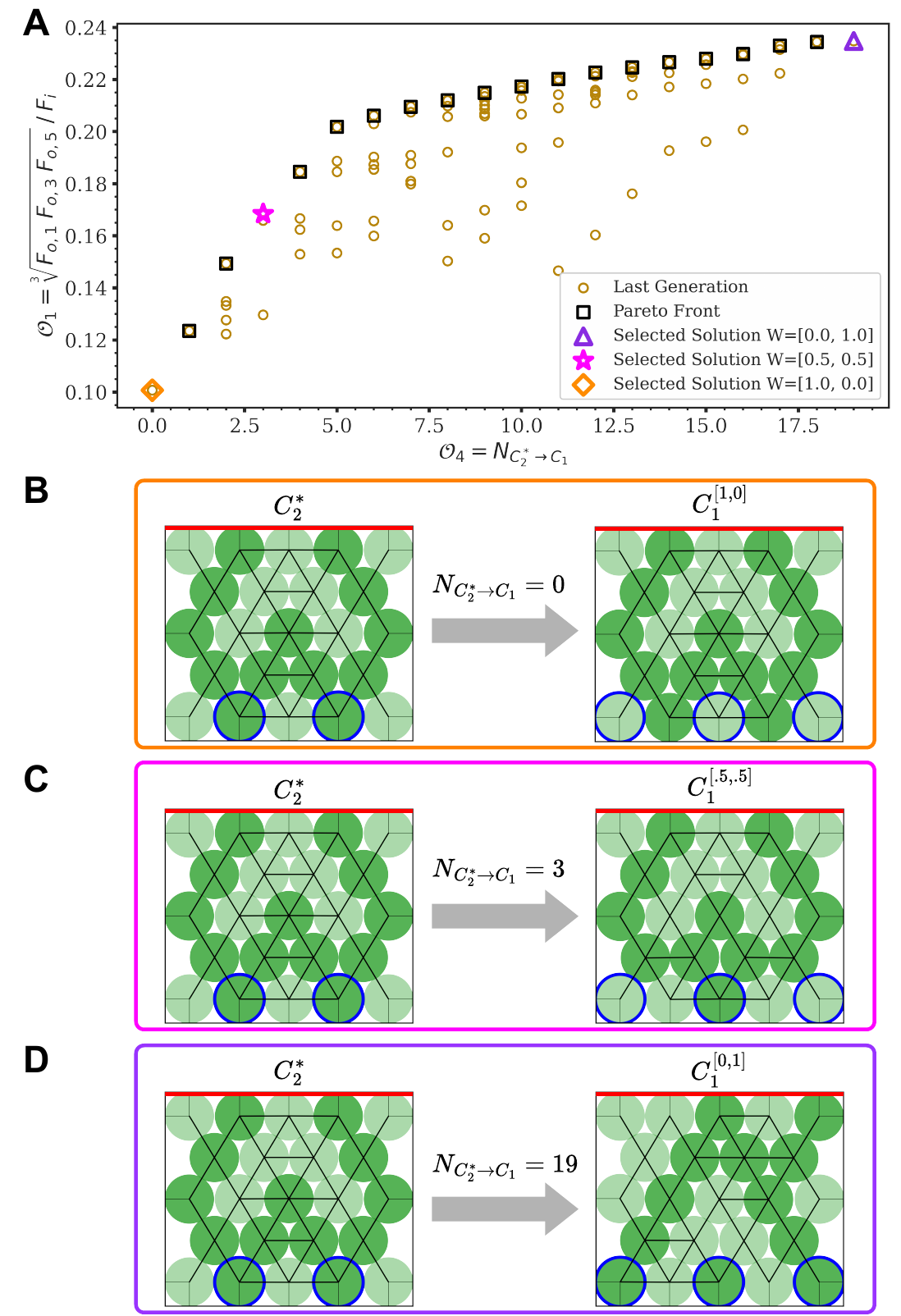}
\caption{Multi-objective optimization to find a configuration that maximizes the output forces on the odd-numbered particles while minimizing the number of particle modulus changes starting from $C_2^*$. (A) Solutions in the $\mathcal{O}_4$-$\mathcal{O}_1$ plane at $G=500$ (yellow circles). We also highlight the Pareto front (black squares). The orange diamond, magenta star, and purple triangle indicate the best solutions based on the TOPSIS method with different weight vectors $W$. (B-D) Starting from $C_2^*$ (left), we show the best solution (right) from the Pareto front with a given TOPSIS weight vector $W$: (B) $W=[1, 0]$, $C^{[1,0]}_1$, (C) $W=[0.5, 0.5]$, $C^{[.5,.5]}_1$, and (D) $W=[0, 1]$, $C^{[0,1]}_1$. The first and second elements of $W$ correspond to the weights for $\mathcal{O}_4$ and $\mathcal{O}_1$, respectively.}\label{fig7}
\end{figure}

We ran the multi-objective optimization using AFPO with parameters outlined in Table~\ref{tbl:parameters}. The algorithm converged to a Pareto front with $N_s+1=20$ optimal solutions at $G=500$, as shown in Fig.~\ref{fig6}A. There is no ``best'' solution among the $20$ solutions on the Pareto front if we do not consider trade-offs between maximizing ${\cal O}_2$ and minimizing ${\cal O}_3$. To implement trade-offs, we include a TOPSIS weight vector: $W = [w, 1 - w]$ with $0 \le w \le 1$. The first and second components of $W$ correspond to the weight for $\mathcal{O}_3$ and $\mathcal{O}_2$, respectively. If we set $w = 1$, the best solution is $C_1^*$ with no particle modulus changes (Fig.~\ref{fig6}B). Similarly, if we set $w = 0$, only objective $\mathcal{O}_2$ is considered, and the best solution is $C_2^*$ (Fig.~\ref{fig6}D). These two TOPSIS weight vectors have obvious solutions, and confirm that $C_1^*$ and $C_2^*$ are on the Pareto front. However, if we set $0 < w < 1$ (\textit{e.g.}, $w = 0.5$), the best solution is not $C_1^*$ or $C_2^*$, and is instead a compromise between maximizing ${\cal O}_3$ and minimizing ${\cal O}_2$, as shown in Fig.~\ref{fig6}C.

For Case IV, we consider the following multi-objective optimization problem. Starting from $C_2^*$, we seek to maximize $\mathcal{O}_1$ and minimize
\begin{equation}
\label{eq7}
    \mathcal{O}_4 = N_{C_2^* \rightarrow C_1} = \sum_{p=1}^N \left(1 - \delta \left(k^{C_2^*, p}, k^{C_1, p} \right) \right).
\end{equation}
We again observe a Pareto front with $N_s+1=20$ optimal solutions at $G=500$, as shown in Fig.~\ref{fig7}A. We implement a similar TOPSIS weight vector: $W = [w, 1 - w]$, with the first and second components corresponding to the weight for $\mathcal{O}_4$ and $\mathcal{O}_1$, respectively. If we set $w = 1$, the best solution is $C_2^*$ with no particle modulus changes (Fig.~\ref{fig7}B). Similarly, if we set $w = 0$, the best solution is $C_1^*$, where only objective $\mathcal{O}_1$ is considered (Fig.~\ref{fig7}D). If we set $0 < w < 1$ (\textit{e.g.}, $w = 0.5$), the best solution is not $C_1^*$ or $C_2^*$, and is instead a compromise between maximizing ${\cal O}_3$ and minimizing ${\cal O}_2$, as shown in Fig.~\ref{fig7}C.

\section{Conclusions}
\label{sec:conclusions}

In this article, we developed a pipeline that combines experiments using variable modulus (VM) particles, discrete element method (DEM) simulations, and evolutionary algorithms (EAs) to design granular packings with adaptive mechanical responses. Specifically, we use the pipeline to design granular packings composed of $N=23$ same-sized VM particles arranged in a triangular lattice with different patterns of particle moduli that maximize the output forces from particles on the bottom row of the system. We realized VM particles in experiments by embedding a Field's metal core within a silicone shell. By running current through a resistive heater co-located with the Field's metal, we achieved a modulus ratio of $k_{stiff}/k_{soft} \sim 3$ when the Field's metal is heated above the melting temperature. The system boundaries were made of photoelastic material to enable measurements of the output forces. We characterized the inter-particle contact forces as a function of compressive strain in experiments to calibrate the inter-particle forces in the DEM simulations. We carried out evolutionary algorithms (EA) to identify the optimal configurations of soft and stiff particles for specific force outputs on the bottom boundary without enumerating all possible particle modulus combinations.

We considered five optimization cases in this study. For the \textit{Feasibility Check}, we found the configuration $C^*$ of particle moduli that maximizes the output force from the middle particle in the bottom row. In \textit{Case I} (\textit{Case II}), we identified the configuration $C_1^*$ ($C_2^*$) that maximizes the output forces from the odd-numbered (even-numbered) particles in the bottom row. In \textit{Case III} (\textit{Case IV}), we found the configuration $C_2$ ($C_1$) that maximizes the output forces from the even-numbered (odd-numbered) particles in the bottom row, while minimizing the number of particle modulus changes starting from $C_1^*$ ($C_2^*$). The first three cases involve single-objective optimization, and the remaining two involve multi-objective optimization. We have shown that EA converges to the optimal solution for the cases involving single-objective optimization and a set of solutions on the Pareto front for the cases involving multi-objective optimization. We verify that the force outputs for the best solutions match in experiments and DEM simulations for the cases of single-objective optimization. We further show that we can identify the best solution in the cases of multi-objective optimization when trade-offs between the objectives are implemented using a TOPSIS weight vector. This study has demonstrated the ability to search efficiently through a large ensemble of granular packings for those that satisfy specific objectives concerning the mechanical response. Although there have been several numerical studies aimed at identifying optimal properties in granular packings using EA~\cite{wu2019active}, few have validated the EA methods using experiments~\cite{miskin2014evolving}. 

We envision several interesting directions for future studies. First, we can examine all of the solutions on the Pareto front in Cases III and IV, implement the optimal solutions in experiments, and determine the power consumption when generating them. Second, our results can serve as a proof-of-concept for primitive mechanical logic gates in granular materials.  Previous work has shown logic capabilities within granular materials \cite{Daraiologic}, but these studies relied on the embedded capabilities of the granular material rather than optimizing for a desired logical gate. Although no traditional form of logic has been investigated in the current study, we believe that the proposed pipeline can serve as a tool to optimize a material to perform logical operations.

The heterogeneity of force chain networks in granular packings is often controlled by positional disorder. Previous studies of force chain networks in granular materials have analyzed force propagation in disordered networks using methods from statistical physics and mathematics~\cite{PersistenceHomology, Edwards2003, RandomMatrix, Homology2, Kondic_2012}.  In this work, we tuned the force chain networks through particle modulus variations ($k_{stiff}/k_{soft} \sim 3$) on a fixed triangular lattice, rather than through positional disorder. Although this modulus ratio is substantial and has allowed us to significantly vary the morphology of force chain networks, an optimized particle design can induce a more dramatic particle modulus ratio and thus more dramatic changes in the force chain networks in VM particle packings. To efficiently sample and quantify force chain networks, we implemented DEM simulations with a simplified contact model for VM particles and the system boundaries. However, the experimental system and the DEM simulations can have differing force outputs for some of the sub-optimal configurations. Increasing the complexity of the inter-particle contact model in the DEM simulations, \textit{e.g.}, using deformable particle simulations in 3D~\cite{wang2021softmatter}, can improve the agreement between the predicted and experimentally measured force outputs.

\section*{Author Contributions}
S.W.: Methodology, validation, formal analysis, investigation, data curation, writing, review and editing, visualization, and project administration. A.P.: Methodology, formal analysis, validation, writing, review and editing, and visualization. D.W.: Methodology, formal analysis, writing, review and editing, and visualization. N.P.: Methodology, investigation, supervision, review, and editing. J.Z.: Methodology, formal analysis. A.K.M.: Methodology. M. D. S.: Investigation, resources. 
R.K.-B.: Conceptualization, supervision, project administration, review and editing, resources, and funding acquisition. C.S.O.: Conceptualization, supervision, project administration, review and editing, resources, and funding acquisition. J.B.: Conceptualization, supervision, project administration, review and editing, resources, and funding acquisition.

\section*{Conflicts of interest}
There are no conflicts to declare.

\section*{Data availability}
We have deposited all relevant data for this paper in the repository \href{https://github.com/AtoosaParsa/AdaptiveForceChains}{https://github.com/AtoosaParsa/AdaptiveForceChains}.

\section*{Acknowledgements}
This material is based upon work supported by the National Science Foundation under the Designing Materials to Revolutionize and Engineer our Future (DMREF) program (award no. 2118988).

\widetext
\clearpage
\begin{center}
\textbf{\large Electronic Supplementary Information (ESI)}
\end{center}
\setcounter{equation}{0}
\setcounter{figure}{0}
\setcounter{table}{0}
\setcounter{section}{0}
\setcounter{page}{1}
\makeatletter
\renewcommand{\theequation}{S\arabic{equation}}
\renewcommand{\thefigure}{S\arabic{figure}}
\renewcommand{\bibnumfmt}[1]{[S#1]}
\renewcommand{\citenumfont}[1]{S#1}
\renewcommand{\thesection}{S\arabic{section}}

\section{Power Requirements}\label{sectS1}

\begin{figure}[!ht]
    \centering
    \includegraphics[width = \linewidth]{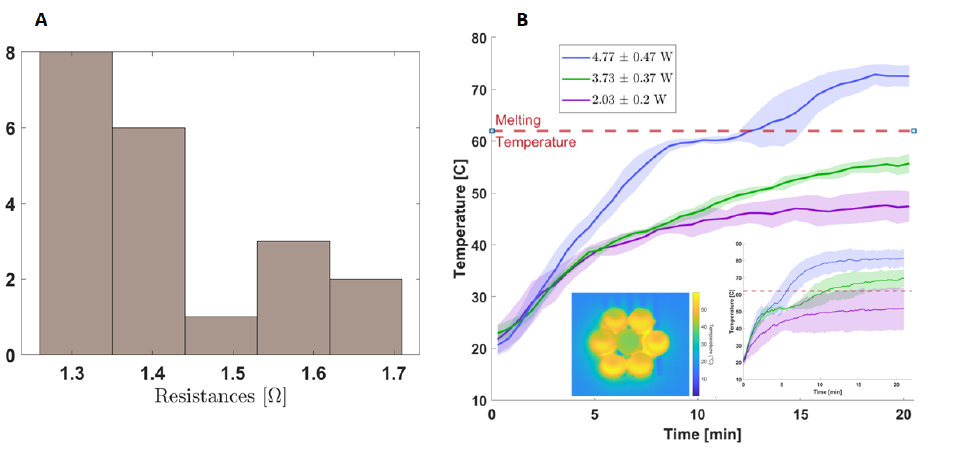}
    \caption{A) Histogram of resistance values for variable modulus particles. B) Temperature as a function of time for a non-actuated particle that is surrounded by actuated particles (inset) for different power inputs.}\label{figs1}
\end{figure}

In actuating the variable modulus (VM) particles, we need to determine the appropriate power setting to use. Using too little power is not enough to melt the actuated particles, whereas too much power will result in significant thermal cross talk, leading to melting of un-actuated particles. We characterize this behavior by placing a single un-actuated particle in the middle of six surrounding actuated particles. Using different power settings, we measure the temperature of the actuated and non-actuated particles as a function of time. The power values are calculated using constant voltage and the measured mean resistances ($P = V^2/R$), as shown in Fig.~\ref{figs1}.

\newpage
\section{Normalization Procedure}\label{sectS2}

\begin{figure}[!ht]
    \centering
    \includegraphics[width = 0.5\linewidth]{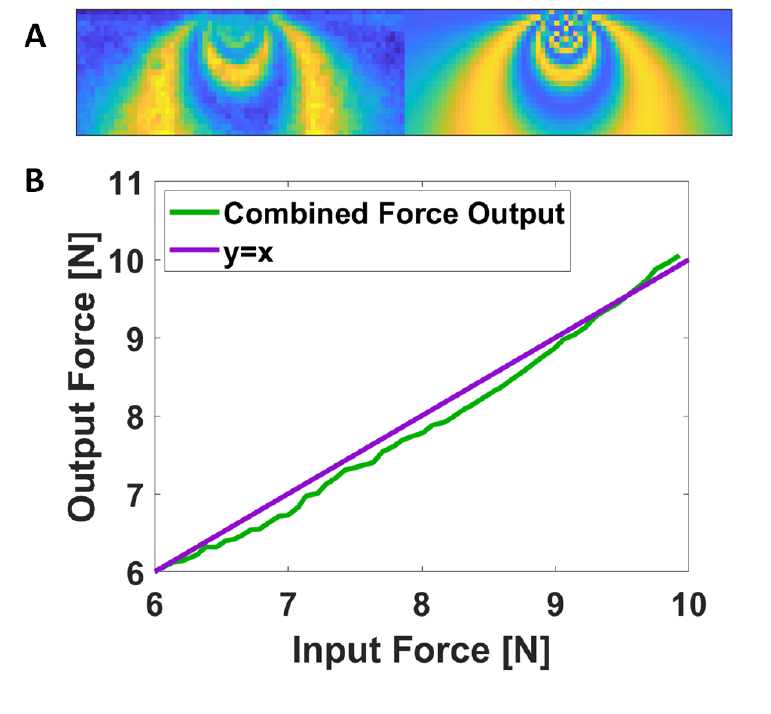}
    \caption{Procedure for obtaining the force on the boundary $F_{o,p}$ from the photoelastic image: A) (left) Photoelastic fringe pattern from the experiments. (right) Photoelastic pattern generated after fitting the output force to Eqs.~5 and 6 in the main text. B) The output  force $F_{o,p}$ on the boundary from a particle on the bottom row plotted versus the input force $R_i$ applied to the top boundary.}\label{figs2}
\end{figure}

We extract the output forces $F_{o,p}$ from the polariscope by chi-square fitting the intensity image (e.g. in Fig.~\ref{figs2}A) using Eqs.~5 and 6 in the main text. This procedure allows us to obtain $2tKF_{o,p}/\lambda$. We can determine the proportionality constant $2tK/\lambda$ by equating the total output force to the total input force we impart the system: $\sum_{p=1}^{5} F_{o,p} = F_i$. The output force on the bottom plate versus the input force is shown in Fig.~\ref{figs2}B.

\newpage

\section{Linear Relationship between $F_{o,p}$ and $F_i$}
\label{sectS:f_in_out}

\begin{figure}[!h]
    \centering
    \includegraphics[width = 0.5\linewidth]{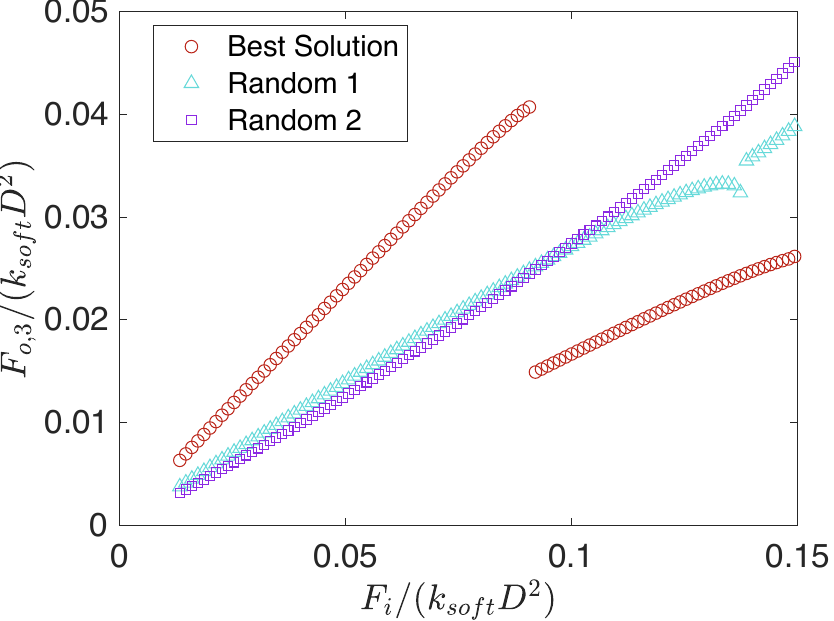}
    \caption{Output force that the third particle in the bottom row exerts on the boundary $F_{o,3}$ plotted versus the input force $F_{i}$ for the three configurations (``Best Solution'', ``Random 1'', and ``Random 2'' in Fig.~4 of the main text). Both $F_i$ and $F_{o,3}$ are normalized by the compressive modulus $k_{soft}D^2$, where $D$ is the diameter of the hard and soft particles.}
    \label{figs:f_in_out}
\end{figure}

We show through DEM simulations that the output force $F_{o,3}$ (\textit{i.e.}, the force that the middle particle in the bottom row exerts on the boundary) is roughly proportional to the input force $F_i$ over a wide range of input forces as long as the interparticle contact network does not change, even though the interparticle force law is non-linear as a function of the particle deformation. In Fig.~\ref{figs:f_in_out}, we plot $F_{o,3}$ versus $F_{i}$ for the three example configurations, the ``Best Solution'', ``Random 1'', and ``Random 2'' in Fig.~4E in the main text, where the forces have been normalized by the compressive modulus of the soft particle $k_{soft}D^2$ and $D$ is the diameter of both stiff and soft particles. We show that $F_{o,3}$ is proportional to $F_i$ for rescaled $F_i/(k_{soft}D^2) \le 0.1$ until a discontinous drop occurs for the ``Best Solution,'' which signals a particle rearrangement. In experiments, the largest input force is $F_i \sim 10$~N, which corresponds to $F_i/(k_{soft}D^2) \sim 0.2$. In the experiments, we do not observe particle rearrangements over the full range of input forces. Particle rearrangements occur in the DEM simulations at lower input forces than in the experiments since the DEM simulations do not include interparticle friction and friction between the particles and the boundaries.

\end{document}